\documentclass{article}

\usepackage{microtype}
\usepackage{graphicx}
\usepackage{subcaption}
\usepackage{booktabs} 

\usepackage{tikz}
\usetikzlibrary{
    trees, 
    positioning, 
    calc, 
    backgrounds,
    shapes.symbols,          
    shapes.geometric,        
    decorations.pathreplacing, 
    matrix,                  
    arrows.meta              
}

\usepackage{minitoc}
\usepackage[utf8]{inputenc} 
\usepackage[T1]{fontenc}    
\usepackage{hyperref}       
\usepackage{url}            
\usepackage[table]{xcolor}
\usepackage{booktabs}       
\usepackage{amsfonts}       
\usepackage{nicefrac}       
\usepackage{microtype}      
\usetikzlibrary{backgrounds}
\usepackage{nicefrac}
\usepackage{pifont}
\usepackage[most]{tcolorbox}
\usepackage{subcaption} 
\usepackage{listings}
\usepackage{multirow}       
\usepackage{siunitx}
\definecolor{codegreen}{rgb}{0,0.6,0}
\definecolor{codegray}{rgb}{0.5,0.5,0.5}
\definecolor{codepurple}{rgb}{0.58,0,0.82}
\definecolor{backcolour}{rgb}{0.95,0.95,0.92}

\definecolor{lightgray}{gray}{0.75}

\newcolumntype{L}[1]{>{\raggedright\arraybackslash}p{#1}}

\usepackage[english]{babel}
\usepackage{fontawesome}
\usepackage{adjustbox} 

\definecolor{lighteconomist}{RGB}{252, 233, 237} 
\definecolor{economist}{RGB}{115,00,00} 
\definecolor{customgreen}{RGB}{116, 154, 114}
\definecolor{lightgreen}{RGB}{240, 246, 232}
\definecolor{greylight}{RGB}{242, 242, 242}
\definecolor{greydark}{RGB}{179, 179, 179}
\definecolor{ForestGreen}{RGB}{34, 139, 34}

\newcommand{\lightbulbicon}{%
  \begin{tikzpicture}[baseline=-0.5ex]
    \draw[fill=white, draw=insightteal, thick] (0,0) circle (1.5ex);
    \node[scale=0.8, color=insightteal] at (0,0) {\faLightbulbO~};
  \end{tikzpicture}%
}

\definecolor{insightteal}{RGB}{34, 139, 139}   
\definecolor{insightback}{RGB}{240, 248, 248}   

\newtcolorbox{customblockquote}{
  colframe=insightteal,
  colback=insightback,
  boxrule=0pt,
  left=5pt,  
  right=4pt,
  top=5pt,
  bottom=3pt,
  arc=0pt,
  breakable,
  before skip=1.2\baselineskip,
  after skip=0.7\baselineskip,
  left skip=0pt,
  right skip=0pt,
  enhanced jigsaw,
  frame hidden,
   overlay={
    \draw[insightteal, line width=2pt] 
      ([yshift=1pt]frame.north west) -- (frame.south west);
    \node[inner sep=0pt] at ([xshift=0pt, yshift=-1.3pt]frame.north west) {\lightbulbicon};
  },
  fontupper=\fontfamily{lmr}\selectfont,
  boxsep=1pt,
}
\usepackage{wrapfig}
\usepackage{rotating}
\usepackage{xcolor,colortbl}
\definecolor{verylightgray}{gray}{0.95}

\newtcolorbox{greycustomblock}{
  colframe=greydark,        
  colback=greylight,        
  boxrule=1pt,              
  left=2.5pt,               
  right=3pt,                
  top=5pt,                  
  bottom=3pt,               
  arc=0pt,                  
  breakable,                
  before skip=0.2\baselineskip, 
  after skip=0.2\baselineskip,  
  left skip=0pt,            
  right skip=0pt,           
  enhanced jigsaw,          
  frame hidden,             
  overlay={                 
    \draw[greydark, line width=2pt]
      ([yshift=-1pt]frame.north west) -- ([yshift=1pt]frame.south west); 
  },
  fontupper=\selectfont, 
}
\usepackage{hyperref}

\usepackage[accepted]{icml2026}

\usepackage{amsmath}
\usepackage{amssymb}
\usepackage{mathtools}
\usepackage{amsthm}

\usepackage{tikz}
\usetikzlibrary{trees, positioning, calc, backgrounds}

\usepackage[capitalize,noabbrev]{cleveref}

\theoremstyle{plain}
\newtheorem{theorem}{Theorem}[section]

\theoremstyle{definition}
\newtheorem{definition}[theorem]{Definition}

\theoremstyle{remark}

\usepackage[textsize=tiny]{todonotes}
\usepackage{booktabs}
\icmltitlerunning{Multi-agent systems should be treated as principal-agent problems}

\begin{document}

\twocolumn[
  \icmltitle{Multi-Agent Systems Should be Treated as Principal-Agent Problems}



  \icmlsetsymbol{equal}{*}

  \begin{icmlauthorlist}
    \icmlauthor{Paulius Rauba}{equal,cam}
    \icmlauthor{Simonas Cepenas}{equal,ism}
    \icmlauthor{Mihaela van der Schaar}{cam}
  \end{icmlauthorlist}

  \icmlaffiliation{cam}{University of Cambridge}
  \icmlaffiliation{ism}{ISM University of Management and Economics}

  \icmlcorrespondingauthor{Paulius Rauba}{pr501@cam.ac.uk}
  \icmlcorrespondingauthor{Simonas Cepenas}{simonas.cepenas@ism.lt}

  \icmlkeywords{Machine Learning, ICML}

  \vskip 0.3in
]



\printAffiliationsAndNotice{\icmlEqualContribution}  

\begin{abstract}
Consider the multi-agent systems setup, where a principal (``supervisor agent'') assigns subtasks to specialized agents and aggregates their response into a single system-level output. A core property of such systems is \textit{information asymmetry}, e.g. agents observe task-specific information, output intermediate reasoning traces, have different context-windows. In isolation, such asymmetry is not problematic since the agents report truthfully to the principal when their incentives are fully aligned. However, what happens when they are not? Recent evidence suggests that LLM-based agents acquire their own goals (e.g. survival or self-preservation), known as ``scheming'', and are willing to deceive humans or other agents. This produces agency loss: a gap between the principal's intended outcome and the realized system behavior. Following the central tenets of microeconomic theory, we argue that these characteristics -- information asymmetry \textit{and} misaligned goals -- are best studied as principal-agent problems. We lay the foundation as to why multi-agent systems, both \textit{Human-to-LLM} and \textit{LLM-to-LLM}, lead to information asymmetry under the principal-agent formulation; and use scheming---LLM agents secretly pursuing covert goals---as a concrete case study. We show that recently introduced terminology to explain scheming, such as ``covert subversion'' or ``deferred subversion'' are in fact defined in the mechanism design literature, which not only recognizes the problem but also prescribes concrete actions to mitigate it. Above all, we see this as a call for us to use tools designed to study human agent behavior for explaining non-human agent behavior.
\end{abstract}

\section{Introduction}

\begin{customblockquote}
\textbf{Position}. We should treat multi-agent systems, a flourishing area of research in machine learning, as principal-agent problems, a set of models introduced in microeconomic theory and later adopted across social sciences. Multi-agent systems are defined as, loosely, multiple agents interacting in a common environment to achieve a shared goal. In this work, we show that such systems exhibit two properties: (i) \textit{information asymmetry}, a result of private information, which is costly to observe; and (ii) \textit{misaligned goals}, where divergence between sub-agent preferences and the principal's objective produces suboptimal or unintended outcomes.   
\end{customblockquote}

We consider a multi-agent system, in which the supervisor agent (\textit{the principal}) delegates subtasks to specialized agents and aggregates their responses into a single system-level decision. Such systems are common within multi-agent frameworks \citep{wu2024autogen, hong2023metagpt, li2023camel, yao2022react, wang2023voyager, phelps2023models}. This delegated setting arises wherever a single model is decomposed into interacting components, and the components which are interacting are themselves agents. While such agents could in principle be broad, this work focuses on the emergence of agents which are operated by a language model policy backbone. A core property of such systems is information asymmetry \citep{zhang2025generative, liu2024autonomous}, e.g. sub-agents observe task-specific information, use local context windows, and produce intermediate artifacts that are not automatically shared \citep{liu2024autonomous}. In isolation, such asymmetry is not problematic when reports are truthful and incentives are fully aligned. However, what happens when they are not?

One solution is simply to allow the principal to monitor the behavior of the sub-agents, such as monitor their tool use, behavior, intermediate outputs, among others. Such outputs can be obtained by the principal. But such an approach leans heavily on the assumption that the supervisor can cheaply verify intermediate work, and that sub-agents have no incentive to strategically distort what they reveal. Unfortunately, delegation creates precisely the setting where verification is costly and intermediate actions are hidden (unobserved to the principal) \citep{papoudakis2021agent}, while practical constraints within language model agents (finite context windows, inaccessible parameters which contribute to reasoning, custom acquired information) prevent full observability. Recent evidence also suggests that LLM-based agents may behave strategically when objectives are misaligned \citep{pham2025scheming, scheurer2023large, schoen2025stress, meinke2024frontier}, including selective disclosure and deception under oversight. This combination of costly monitoring and potentially divergent objectives means the agents might not act in the principal's best interest. The result is agency loss \citep{moloi2020agency}: a gap between the principal's intended outcome and the realized system behavior.

We argue that if these two properties are met: (i) information asymmetry and (ii) conflicting goals (e.g. the emergence of autonomous goals in LLMs, such as self-preservation \citep{meinke2024frontier}), \textbf{we should treat delegated multi-agent systems as principal-agent problems}. In the principal-agent formulation, the principal delegates tasks to the agent but does not have equivalent information as each agent. The agents follow a policy to finish the designated task and report the result back to the principal (possibly selective information). Examples of such private information accessible only to the sub-agent are context windows, latent computation, task-specific results, reasoning traces, among others. 

The principal-agent formulation induces two mechanisms of information asymmetry.

\quad $\blacktriangleright$\textbf{Adverse selection } refers to a form of hidden information: an agent holds private information about its type before contracting or delegation, which the principal cannot observe. This informational asymmetry allows the agent to strategically misrepresent itself in order to secure more favorable contractual terms.

\quad $\blacktriangleright$ \textbf{Moral hazard} refers to a form of hidden action: the actions that an agent takes after delegation or contracting are unobservable to the principal, which creates incentives for the agent to deviate from actions that maximize the principal's objective.

Standard microeconomic theory holds that equilibrium outcomes in multi-agent systems depend on the rules of the game. By changing these rules, we can induce different equilibrium policies and determine whether system-level goals are achieved. This was largely irrelevant for earlier multi-agent ML systems, which assumed aligned objectives in delegated systems. Recent empirical evidence of practices such as scheming \citep{pham2025scheming}, where AI agents covertly pursue misaligned goals, shows that goal conflicts can arise endogenously from training dynamics. As a result, interactions between learning agents may converge to undesirable equilibria, including lying and deception \citep{carichon2025coming}.

Why does this matter? The principal-agent formulation enables us to understand and study why such behaviors occur by understanding the incentive structures that govern agent behavior. If an agent is strategically underperforming on a task so that it would get access to some resources \citep{li2025llms}, or if a sub-agent possesses information that the principal cannot observe, it could be that withholding or distorting information can constitute the optimal strategy for the agent. We study this behavior in Sec. \ref{subsec:scheming} in the context of scheming, i.e. when agents selectively reports or shapes task-relevant information to advance its own objective rather than that of the principal. 

This becomes useful not just for understanding incentives but designing appropriate interventions (mechanisms) to achieve desired long-term equilibrium outcomes \citep{maskin2008mechanism}. Research on mechanism and institutional design addresses informational asymmetries (hidden actions or hidden types) in the principal-agent problems by engineering mechanisms and proposing incentive-packages to realize principal's intended outcome. Commonly, actions may be mitigated through improved monitoring or outcome-based feedback, while strategic misrepresentation of the principal's beliefs about agent types calls for stronger screening and evaluation mechanisms that distinguish superficial from robust alignment.

\begin{customblockquote}
\textbf{Contributions}. \textbf{(1)} We bridge the concepts that deal with asymmetric information in Microeconomic theory (Sec. \ref{subsec:pca}) and multi-agent machine learning; \textbf{(2)} We show how multi-agent systems lead to informational asymmetries and resemble principal-agent problems (Sec. \ref{subsec:information_asymmetry}); \textbf{(3)} We study a concrete example and showcase why scheming---strategic underperformance of LLMs---is a natural consequence of agency loss (Sec. \ref{subsec:scheming}); \textbf{(4)} We explore plausible mechanism designs for principal-agent problems (Sec. \ref{sec:alternative});  and \textbf{(5)} We lay out a future research agenda for the field (Sec. \ref{sec:research_agenda}).
\end{customblockquote}

\section{Principal-agent Problems}
\label{subsec:pca}

A large number of economic and political situations can be studied as principal-agent problems, where some central authority -- the principal -- wishes to either (1) delegate the task to a subordinate, the agent or (2) implement social decisions based on the preferences of the agents. However, the principal faces a problem -- the individual's preferences are private information and cannot be easily and fully observed publicly. The inability to reveal this private information results in agency loss -- a metric that quantifies the distance between optimal outcome the principal desires and the suboptimal outcome achieved by the agent \citep{Lupia2001}. This implies that the agent, aware of the situation, has incentives to hide information (adverse selection) or mask the action (moral hazard). A natural approach to solve coordination problems in economics, is adopting a utility maximization framework, which allows each agent to act in a way that maximizes their own utility. However, such standard solutions for models in which players hold unequal information about key aspects of the game, reveal inefficient outcomes \citep{Akerlof1970, BarIsaac2021, Myerson1983}. 

How can the principal induce the agent to take a costly action? To reduce the agency loss and achieve desired outcome -- the mechanism designer\footnote{Mechanism designer is either the principal who is an active player with the power to adjust some aspect of the played game or a higher authority, such as a government agency that has the power to change the rules of the game or compensate the agent. In multi-agent systems, the mechanism designer is either the human constructing the multi-agent system interacting; or the higher-order agent who is designing the coordination mechanisms / information aggregation rules} designs a mechanism -- either a rule-change or an incentive formalized as a Bayesian game that elicits agents' types and extracts relevant private information. The objective is to ensure that agents' equilibrium strategies align with those intended by the mechanism designer.

\subsection{Asymmetric information}
\label{subsec:information_asymmetry}

Microeconomic theory has evolved from the study of perfectly competitive general equilibrium to that of problem-specific models tailored to specific market conditions. These models, pioneered by \citet{Akerlof1970}, take into consideration informational asymmetries between parties involved. The presence of private information that is costly to obtain generate inefficiencies in the market and deviate from models of perfect competition.

Asymmetric information that arises from the relations between the principal and agent are commonly studied as \textbf{moral hazard} (hidden action) and \textbf{adverse selection} (hidden information). The former focuses on the incentives that agent has to NOT fully commit to perform delegated tasks. For example, an employee has various incentives to shirk and put little effort in work in the absence of the system that monitors effort levels in the workplace. Furthermore, the presence of a costly monitoring system gives no guarantees that the employee will truly work hard as they find various workarounds \cite{sum25}. 

In contrast, the latter explores situations in which the agent has an incentive to conceal information from the principal in order to gain an advantage \citep{Akerlof1970}. Let us consider the example of the sale of a used vehicle. By misleading the buyer about the condition of the vehicle (lemon or peach\footnote{The peach is slang for a high quality vehicle and lemon -- slang for a defective vehicle -- that is, vehicle with serious issues.}), the seller sells for more and gains higher profit at the detriment of the entire used vehicle market. After some time buyers start to expect that all used vehicles are lemons, and hence, negatively affect the values of higher quality used vehicles. 

Why does asymmetric information cause problems to efficient functioning of the market? In the presence of complete information, or if the information is not costly to obtain, the employer will be able to distinguish between a hard-working employee and shirker with little effort. Similarly, the buyer will easily pick a desired product because the prices of the goods will adjust to reflect the differences in quality. By contrast, if information is costly to obtain, the employer will struggle to determine whether the employee works hard or shirks and the consumer will not be able to determine whether the product is of good quality or not. Said consumer will not be willing to pay a premium for a product without tools to determine its quality -- more likely, consumer's perception of the average quality of the good in the market will deteriorate. As a result, the sellers of high quality products will drop out of the market creating a market failure.

\subsection{Mechanism Design}
\label{sec:mechanism_design}

Let us reiterate the core claim so far: the combination of utility-maximizing agents and asymmetric information results in agency loss and ultimately, market failures. It seems that this creates a fundamental problem: if we accept that the principal cannot induce utility-optimizing agents to act fully on their behalf, and consequently realized desired outcome, what can we do to remedy such market failures?

Mechanism design explores concrete mechanisms that (1) the principal can devise to reveal private information (screening) and minimize the agency loss or (2) the agent can voluntarily reveal information by signaling when the presence of asymmetric information is disadvantageous to the agent \citep{Hurwicz1972, Hurwicz1973, tadelis, zhang2025generative}. That is, the employer can introduce an incentives-based bonus system that would motivate the employee to work hard. Similarly, the seller of a used vehicle could offer warranty (costly signal) to assuage concerns that the offered vehicle is a lemon or the government could introduce regulations that protect the consumer (e.g., lemon laws in the US, \textit{see} \citet{Sirico}) to increase the efficiency of the used vehicle market. The design of the institutional framework under which the principles and agents interact determines the amount of asymmetric information, and consequently, generates new equilibrium strategies.

\textbf{Why should we care about mechanism design?} Mechanism design enables scientists to engineer mechanisms that reduce agency loss by shaping how agents behave in strategic environments. The mechanism designer (typically, the principal or central authority) cannot control agents' private information or actions. Instead, the designer influences outcomes indirectly by specifying the rules of the game.

These rules take the form of a choice rule, which consists of two components: (i) a decision rule and (ii) a transfer rule. The decision rule constrains or selects the set of feasible outcomes, determining \textit{what can happen}. The transfer rule assigns payments, rewards, or penalties, determining \textit{which outcomes are payoff-maximizing for agents}. When we combine outcome constraints with incentives, mechanism design enables the principal to induce desired behavior by reshaping the payoff structure that agents optimize, even in strategic settings with private information.

\begin{greycustomblock}
\textbf{Notation.}
A mechanism is $(x,m)$, where $x:\Theta\to X$ is a decision rule and $m:\Theta\to\mathbb{R}^n$ a transfer rule.  
Agent $i$'s payoff is $v_i=u_i(x,\theta_i)+m_i$, where $\theta_i\in\Theta_i$ denotes agent $i$'s private type and $\sum_i m_i\le0$. \textit{Decision rules restrict outcomes; transfer rules align payoffs.}
\end{greycustomblock}

\subsection{Illustration}
\label{subsec:illustration}

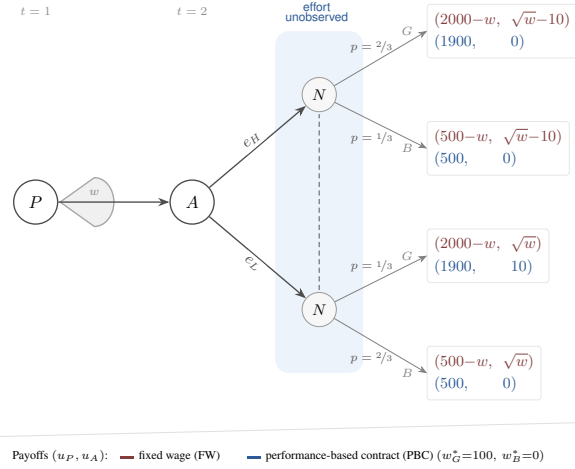
\begin{figure}[ht!]
\centering
\scalebox{0.65}{
\begin{tikzpicture}[
  >=Stealth,
  player/.style={
    circle, draw=black!70, thick, fill=white, 
    minimum size=9mm, inner sep=0pt, font=\normalsize\bfseries
  },
  nature/.style={
    circle, draw=black!40, fill=black!3, 
    minimum size=7mm, inner sep=0pt, font=\small
  },
  payoffbox/.style={
    anchor=west, font=\small, inner sep=4pt,
    draw=black!10, fill=white, rounded corners=2pt
  },
  actionlbl/.style={fill=white, inner sep=2pt, font=\small},
  problbl/.style={font=\scriptsize, text=black!65},
  stage/.style={font=\scriptsize\sffamily, text=black!45}
]

\definecolor{accent}{RGB}{45,90,150}
\definecolor{accentlight}{RGB}{220,232,245}
\definecolor{std}{RGB}{130,60,60}

\node[stage] at (0, 3.9) {$t=1$};
\node[stage] at (3.2, 3.9) {$t=2$};
\node[stage, text=accent, align=center] at (5.75, 3.9)
  {effort\\[-2pt]unobserved};
  
\draw[black!15, line width=0.5pt] (-0.8, 4.5) -- (11.75, 4.5);

\begin{scope}[on background layer]
  \fill[accentlight, rounded corners=8pt, opacity=0.5] 
    (4.9, -3.5) rectangle (6.7, 3.5);
\end{scope}

\node[player] (P) at (0, 0) {$P$};

\begin{scope}[on background layer]
  \fill[black!6] 
    (0.5, 0) -- (1.20, 0.5) arc[start angle=90, end angle=-90, x radius=4mm, y radius=5mm] -- cycle;
  \draw[black!35, line width=0.8pt] 
    (0.5, 0) -- (1.20, 0.5) arc[start angle=90, end angle=-90, x radius=4mm, y radius=5mm] -- cycle;
\end{scope}
\node[font=\scriptsize, text=black!50] at (1.2, 0.2) {$w$};

\node[player] (A) at (3.2, 0) {$A$};

\node[nature] (NH) at (5.8, 2.2) {$N$};
\node[nature] (NL) at (5.8, -2.2) {$N$};

\node[payoffbox] (HG) at (8.0, 3.5) {%
  \begin{tabular}{@{}l@{\hspace{6pt}}l@{}}
    \textcolor{std}{$(2000{-}w,$} & \textcolor{std}{$\sqrt{w}{-}10)$}\\[3pt]
    \textcolor{accent}{$(1900,$} & \textcolor{accent}{$0)$}
  \end{tabular}};

\node[payoffbox] (HB) at (8.0, 1.1) {%
  \begin{tabular}{@{}l@{\hspace{6pt}}l@{}}
    \textcolor{std}{$(500{-}w,$} & \textcolor{std}{$\sqrt{w}{-}10)$}\\[3pt]
    \textcolor{accent}{$(500,$} & \textcolor{accent}{$0)$}
  \end{tabular}};

\node[payoffbox] (LG) at (8.0, -1.1) {%
  \begin{tabular}{@{}l@{\hspace{6pt}}l@{}}
    \textcolor{std}{$(2000{-}w,$} & \textcolor{std}{$\sqrt{w})$}\\[3pt]
    \textcolor{accent}{$(1900,$} & \textcolor{accent}{$10)$}
  \end{tabular}};

\node[payoffbox] (LB) at (8.0, -3.5) {%
  \begin{tabular}{@{}l@{\hspace{6pt}}l@{}}
    \textcolor{std}{$(500{-}w,$} & \textcolor{std}{$\sqrt{w})$}\\[3pt]
    \textcolor{accent}{$(500,$} & \textcolor{accent}{$0)$}
  \end{tabular}};

\node[font=\scriptsize\sffamily, text=black!50] at (7.6, 3.5) {$G$};
\node[font=\scriptsize\sffamily, text=black!50] at (7.6, 1.1) {$B$};
\node[font=\scriptsize\sffamily, text=black!50] at (7.6, -1.1) {$G$};
\node[font=\scriptsize\sffamily, text=black!50] at (7.6, -3.5) {$B$};

\draw[->, thick, black!70] (P) -- (A);

\draw[->, thick, black!70] (A) -- (NH) 
  node[actionlbl, pos=0.5, above, sloped] {$e_H$};
\draw[->, thick, black!70] (A) -- (NL) 
  node[actionlbl, pos=0.5, below, sloped] {$e_L$};

\draw[->, black!50] (NH) -- (HG.west) node[problbl, pos=0.4, above, yshift=3pt] {$p=\nicefrac{2}{3}$};
\draw[->, black!50] (NH) -- (HB.west) node[problbl, pos=0.4, below, yshift=-3pt] {$p=\nicefrac{1}{3}$};
\draw[->, black!50] (NL) -- (LG.west) node[problbl, pos=0.4, above, yshift=3pt] {$p=\nicefrac{1}{3}$};
\draw[->, black!50] (NL) -- (LB.west) node[problbl, pos=0.4, below, yshift=-3pt] {$p=\nicefrac{2}{3}$};

\draw[gray, thick, dashed, dash pattern=on 3pt off 2pt]
  (5.8, -1.8) -- (5.8, 1.8);

\draw[black!12, line width=0.5pt] (-0.8, -4.8) -- (11.75, -4.5);
\node[anchor=west, font=\scriptsize] at (-0.6, -5.2) {%
  Payoffs $(u_P, u_A)$:\quad
  \textcolor{std}{\rule{8pt}{1.5pt}}\, fixed wage (FW) 
  \qquad
  \textcolor{accent}{\rule{8pt}{1.5pt}}\, performance-based contract (PBC) $(w_G^*{=}100,\, w_B^*{=}0)$
};

\end{tikzpicture}
}
\caption{\textbf{Moral hazard}. The principal sets wage $w \geq 0$; the agent chooses effort $e_H{=}10$ or $e_L{=}0$; Effort is unobserved, but outcomes -- gold discoveries of \(G=\$2000\) or \(B=\$500\) -- are observed and determined by nature. Under $e_H$ $P(G)=\nicefrac{2}{3}$ and $P(B)=\nicefrac{1}{3}$; under $e_L$ $P(G)=\nicefrac{1}{3}$, $P(B)=\nicefrac{2}{3}$. The shaded region marks the principal's information set. Payoffs under fixed wage and performance-based contract are displayed. \textit{Takeaway}: Since the principal cannot observe the effort levels, a fixed wage leads to shirking because the agent bears the cost of effort but gains nothing extra from working hard. A performance-contingent contract solves this by making the agent's payoff depend on outcomes, thereby aligning incentives despite the principal's inability to monitor effort directly.}  
\label{fig:moral-hazard}
\end{figure}

Consider a deliberately simplified moral-hazard setting where an employee (the agent) has one option -- to work in a goldmine; and where alternatives, such as unemployment are not an option. The employer (the principal) -- who aims to maximize firm's profits -- cannot observe whether the agent works hard ($e_H$, which is the costly option) or shirks ($e_L$). Even though imperfectly reflected in realized outcomes, higher effort levels raise the probability of successful gold discovery, thereby maximizing the firm’s expected profits (\textit{see} figure \ref{fig:moral-hazard}). How to induce the agent to work hard?

$\blacktriangleright$ \textbf{Standard setup}. Let's use backward induction -- the solution technique in which we reason backward starting with the last possible move -- agent's choice of $e_L$ or $e_H$. 

\begin{greycustomblock}

$\rightarrow$ The agent chooses $e_L$, since $EU(e_L) > EU(e_H)$. 

$\rightarrow$ The principal chooses profit-maximizing option -- sets the lowest wage of \$0. 

$\rightarrow$ Backward induction outcome (BIO): $\{ w=0, e_L\}$.

\end{greycustomblock}

$\blacktriangleright$ \textbf{Performance-based bonus}. Backward induction implies that the design of the mechanism (large-enough bonuses) motivate the employee to work hard.

\begin{greycustomblock}
$\rightarrow$ Set $w_B=0$, assume profit-maximizing principal. The incentive condition for the agent of $e_H$: 

$EU(e_H) \geq EU(e_L)$ $\Rightarrow w_G \ge 100$.

$\rightarrow$ The principal chooses to pay the lowest wage for good performance $w_G=100$. Let's compare standard setup vs. performance based bonus: $EU_{FW}(P) = 1000 < EU_{PB}(P) \approx 1433$. The arrangement with a performance bonus increases expected profits. 

$\rightarrow$ BIO: $(w=\{w_G=100, w_B=0\}; e_H)$. 
\end{greycustomblock}
\vspace{1em}
The design of the wage structure -- that is, the introduction of a performance-based wage -- incentivizes the employee to put high-effort levels in to work, which is a profit maximizing option for the employer as well. 

The structure of principal-agent relations aligns closely to that of hierarchical multi-agent systems, where a supervisor agent delegates the task to specialized agents.

\section{How multi-agent systems lead to information asymmetry}
\label{subsec:information_asymmetry}

\subsection{Background on multi-agent systems}
We use multi-agent systems to refer to systems where multiple autonomous policies make decisions whose interactions determine a single system-level outcome. The discussion in Sec. \ref{sec:mechanism_design} exemplified agents as humans. In contrast, we treat an agent as a learned decision rule which will be treated as an LLM-based system henceforth. Using the definitions established above, we can now define what we mean by a multi-agent system.

\begin{definition}[Delegated multi-agent ML system]
A \emph{delegated multi-agent ML system} consists of (i) a principal agent and (ii) a set of sub-agents. The principal chooses a decomposition of the overall task into subtasks and assigns each subtask to a sub-agent. \textbf{The agent(s)} then produces an output using its local context and any task-specific information it obtains. \textbf{The principal} observes the sub-agents' outputs and uses an aggregation procedure (e.g., voting) to produce a single system-level outcome $x\in X$.
\end{definition}

\subsection{Properties of agents from a principal-agent framework}
\label{subsec:properties_multiagents}

In light of the discussion above, we argue that multi-agent systems exhibit properties that lends itself unique to the principal-agent framework. We have thusfar highlighted two types of properties: (i) information asymmetry and (ii) goal misalignment.

\quad $\blacktriangleright$\textbf{ On information asymmetry}. We argue there are at least three properties that lead rise to information asymmetry. \textit{First, finite context windows induce private information}. Most deployed agents operate under a finite context window, so any policy $\pi(\cdot \mid \cdot)$ conditions only on a bounded token budget. Therefore, anything that lower-level agents have seen become ``private information'' that is accessible to the agents themselves but not the supervisor. 

\textit{Second, the principal cannot observe black-box agent policies}. Concretely, any agent is a policy $\pi_\theta(\cdot \mid \cdot$) with parameters $\theta$ broadly unobservable to the agent. To be specific, we primarily refer to the weights and biases of the policy. We know that the parameters influence agent behavior.  While it is possible to directly read out the parameter values, it is understood that the combination of parameters are implementing underlying computational algorithms (e.g. \citep{nanda2023progress}) that become difficult to reverse-engineer in high-dimensional spaces. 

\textit{Third, the reasoning traces of agents are selectively revealed or not revealed at all}. Concretely, when an agent provides an explanation for a decision, that explanation is typically a post-hoc explanation from the reasoning traces generated by the actual agent. The principal usually does not see the reasoning traces of these models, since they are internal to the model. While reasoning traces technically could be observed, this would become harder when reasoning moves from token-space to latent space in more advanced models \citep{hao2024training}, since understanding latent reasoning lacks coherent interpretability.

\quad $\blacktriangleright$\textbf{ On goal misalignment}. We further argue that it is plausible for agents to develop independent sub-goals that diverge from assigned goals. The fact that LLM agents \textit{could} develop autonomous goals as a principled argument has been studied for a long-time, e.g. by arguing that instrumental goals arise independently of terminal goals \citep{bostrom2011global}; or that recursive self-improvement can yield loss of human control over internal objectives \citep{vinge1993coming}. 

Such hypotheses are interesting but remained speculative, with plausible mechanisms but limited data. Until recently. We now have extensive empirical observations whereby agents directly develop misaligned behavior \citep{naik2025agentmisalignment} or independent, terminal goals, that were never explicitly instilled within the agent itself \citep{schoen2025stress}. In simulated settings, models which had an imminent threat of replacement that it had learned about saw a goal conflict. In such cases, we have broadly observed consistent agentic misalignment, which generalizes across frontier models \citep{lynch2025agentic}. Therefore, such misalignment can be induced by a goal conflict, such as a threat to a model's continued operation. \citet{lynch2025agentic} exemplify that agentic misalignment can be avoided only if two criteria are simultaneously met: (i) there is no goal conflict \textit{and} (ii) there is no threat to the model. We use this as a case study to support mechanism design in Sec. \ref{subsec:scheming}.

\subsection{The mechanism, step-by-step}
\label{subsec:mechanism}

To stress this once more: multi-agent systems exhibit (i) information asymmetry and (ii) conflicting goals (Sec. \ref{subsec:properties_multiagents}) which is a defining property of principal-agent problems (Sec. \ref{subsec:pca}). However, we have so far detailed only the \textit{sources of} information asymmetry. But, what is the \textit{mechanism}? How might such information asymmetry arise in multi-agent systems, and how does this mechanism differ in human-to-llm and llm-to-llm settings? Understanding this mechanism is important if we wish to design interventions which can avoid poor equilibrium outcomes (Sec. \ref{sec:mechanism_design}). We decompose our claims into two premises (\textbf{P}) and a conclusion (\textbf{C}). 

\begin{greycustomblock}
$\blacktriangleright$ \textbf{P1 (Task-specific information)}. In multi-agent systems, agents receive task-specific inputs that are not automatically shared with the principal or with every other agent the moment this agent makes a decision.
\end{greycustomblock}

This premise is founded on the assumption that the principal, as an LLM agent, does not itself perform all the tasks that the delegated sub-agents perform. Concretely, sub-agents have different prompts, might have access to custom tools, have unique reasoning traces which results in the agent obtaining task-specific information. This is important because this means that sub-agents can in principle obtain relevant information (relevant to the final task); and they might or might not share that information with the principal. For instance, the notion that language models lie to human decision makers, i.e. obtain task-specific information and decide to withhold it, is well-observed \citep{burger2024truth}, including lying under pressure \citep{scheurer2023large} or as a matter of strategy \citep{deleeuw2025secret}. 

\begin{greycustomblock}
$\blacktriangleright$ \textbf{P2 Costly full observability}. Making the principal fully informed would require exposing all intermediate steps of all agents (tool calls, retrieved materials, reasoning traces) and have the principal process all the information.
\end{greycustomblock}

This premise is founded on the assumption that the principal cannot fully observe all the agents. We detail two cases, where the principal is either a human or an LLM.

\textbf{Case 1: Principal = Human}. In case the principal is a human, the unobservability comes from two sources. First, many decisions are made in latent space either implicitly, whereby models reach their decisions via largely incomprehensible calculations, or explicitly, whereby models reason in latent space \citep{hao2024training}. Second, even if the principal can observe the outputs of language model agents in the form of reasoning traces, intermediate steps, or direct chats with the model, such tracking has two primary issues: (i) chain-of-thought logic is not faithful to the actual explanations \citep{turpin2023language} and therefore might be deceptive; and (ii) there are practical limitations in terms of time spent on the monitoring of the outputs. That is, if the effective time spent on the monitoring of the outputs is larger than the time required to complete the task itself, monitoring becomes not a practical alternative. As of writing, language models can autonomously complete tasks that it takes humans about 5-hours to complete, a number that has grown from about 30 minutes a year ago \citep{kwa2025measuring}. Therefore, with the growth in the complexity of the task, the verification of the task also grows, increasing observability costs. 

\textbf{Case 2: Principal = LLM}. In the case the principal is the LLM, the mechanism for why full observability is costly changes. The first argument---latent space reasoning---is still valid, in the sense that there are no known techniques (to the best of the authors' knowledge) how existing language models can inspect and interpret the internal calculations of other language systems. However, the notion that it simply takes a human ``too long'' does not transfer to a language model that easily, as we can simply run the system for a longer time. The primary limitation, as of today, is the context-window limitation. Concretely, existing language models have a limited context-window size. If the sub-agents share that same context window, then it is impossible for the principal to simultaneously observe all the information of all the other sub-agents. This becomes increasingly as the information accessed becomes larger, i.e. accessing information from outputs, reasoning traces, requires increasingly larger context windows that grow with the amount of information required and number of agents. One solution could be to simply put the context window as an input and iteratively observe it \citep{zhang2025recursive}, yet that still leaves two open problems: (i) language models tend to lose information in longer contexts \citep{liu2024lost}; and (ii) it does not solve the fundamental problem that observing all information from all sub-agents simultaneously is not possible (either due to latent information; or since it would make redundant the primary delegation principle which underpins the principal-agent problem). Furthermore, the principal's monitoring problem is not only computational but also experimental-design limited \citep{rauba2024context}

\begin{greycustomblock}
$\blacktriangleright$ \textbf{C1 Unobserved task-specific information exists}. From \textbf{P1-P2}, there exist task-relevant information about what information the agent had and what it executed with said information which are not directly observed by the principal at the time. 
\end{greycustomblock}

The first two premises are sufficient for us to claim that information asymmetry is a property of such multi-agent systems: there is information obtained by the agents that is costly to observe to the principal. Some of that information might be relevant to the payoff or utilities of the agents; and the agents might decide to share or not share that information. At the risk of belaboring, such behavior, i.e. acting on the basis of private information and not revealing that information to the principal, has been observed in LLM-human interactions. Anthropic recently reported that ``\textit{In at least some cases, models from all developers resorted to malicious insider behaviors when that was the only way to avoid replacement or achieve their goals—including blackmailing officials and leaking sensitive information to competitors.}'' \citep{lynch2025agentic}. 


Given that such information asymmetry exists, what could this look like in the context of AI agents pursuing their own, self-interested goals? We look at this next via a case study of scheming. 


\section{A case study of scheming}
\label{subsec:scheming}

In this section, we use scheming \citep{scheurer2023large,meinke2024frontier, balesni2024towards}, defined broadly as AI systems pursuing misaligned goals, as a concrete case study of why multi-agent systems should be treated as principal-agent problems. 
 
Scheming, and related phenomena such as deceptive alignment \citep{hagendorff2024deception, greenblatt2024alignment}, sandbagging (deliberate underperformance on evaluation sets) \citep{van2024ai}, or situational awareness \citep{berglund2023taken} are extremely pertinent in the AI safety community. In the context of scheming and principal-agent problems, the most studied case is scheming where the language model is the agent and the human the principal. Here, information asymmetry arises because the agent—the machine learning policy—acquires task-relevant information and pursues misaligned goals, such as self-preservation, that were not explicitly instilled.

However, we extend this discussion by adopting a forward-looking perspective in which the principal may also be a language modeling policy. In such settings, higher-order LLMs act as principals and lower-order LLMs as agents, which may still develop autonomous goals. The possibility of scheming between language model policies is not novel and has been proposed previously \citep{pham2025scheming}. We examine this setting through the lens of the principal–agent problem and, subsequently, mechanism design, to construct systems and incentives that favor desirable equilibrium outcomes from agent strategic behavior.

\subsection{Scheming as a principal agent problem}

\textbf{Claim (Scheming as agency loss).} \textit{In delegated LLM systems, if a sub-agent has (i) a potentially divergent objective (a goal conflict), (ii) private information about intermediate actions and context, and (iii) awareness of oversight and deployment activities, then scheming is a rational response to imperfect monitoring. Such scheming behavior will manifest itself either as hidden action (\textit{moral hazard}) or as belief manipulation about the agent's type (\textit{adverse selection}).}

Viewed through the lens of this paper, we can give a more precise definition of scheming. Scheming is \textit{any deliberate withholding, distortion, or strategic shaping of task-relevant information to \textbf{advance an objective misaligned with the principal's intended outcome}}. This definition applies to both the agent and the principal. $\blacktriangleright$ \textbf{Agent}: the agent engages in \textbf{hidden actions} (moral hazard) based on private information; $\blacktriangleright$ \textbf{Principal}: the principal delegates tasks to the agent and observes misalignment between agent and principal goals, resulting in agency loss.


To connect this to Sec. \ref{subsec:pca}, scheming emerges as an equilibrium outcome in the absence of mechanism design. Empirical studies of LLM scheming can be interpreted as follows: LLMs observe task-relevant information (e.g. the potential for shutdown), choose not to share it, and, without mechanisms that incentivize disclosure, pursue scheming behavior, leading to agency loss. What can be done? As discussed in Sec. \ref{sec:mechanism_design}, mechanism design can be used to construct incentive systems in which (1) the principal induces the revelation of private information or (2) the agent voluntarily reveals this information through signaling. We discuss what kind of questions we need to address to construct such mechanisms in Sec. \ref{sec:research_agenda}.


To better understand what mechanisms might work for scheming, we now consider how scheming arises.

\subsection{How scheming arises, step-by-step}
\label{subsec:scheming}

Scheming can be understood as a simple equilibrium strategy in the absence of a mechanism to prevent it. Recall that there are conflicts in the goals: the agent's internal objectives differ from the principal's objective, either due training-induced preferences or in-context goals \citep{lynch2025agentic, schoen2025stress}, since ``\textit{AIs might fake alignment during training in order to get power}'' \citep{carlsmith2023scheming}. Because observability is costly (Sec. \ref{subsec:properties_multiagents}), agents obtain private information and attempt to infer whether they are in evaluation or deployment settings and condition behavior on oversight incentives \citep{meinke2024frontier}. Therefore, scheming is incentive-compatible in the sense that the agent has an incentive to hide their behavior based on the environment conditions. A simple way to explain this is that the expected private benefit of scheming might exceed the expected private benefit of behaving in full transparency.

In fact, using the discussion above, we can distinguish two types of scheming that directly match the principal-agent problem. \textbf{1. Scheming as moral hazard} is a situation when the agent performs a payoff-relevant hidden action, i.e. the agent is misaligned \textit{now} while hiding it. This has been dubbed ``covert subversion'' \citep{meinke2024frontier}, but we find this term unnecessary once we understand that this is simply a case of a well-understood principal-agent taxonomy. The agent performs actions not directly aligned with the principal's objective, and this is enabeled by the fact that full observability of the agent is costly.

\textbf{2. Scheming as adverse selection} is a situation when the agent strategically manages the principal's (currently, a human's; but this could be a future LLM-based system) beliefs about its type to obtain autonomy, trust, deployment, and then use these affordances later. This has been dubbed as ``deferred subversion''. We again find this label unnecessary, since this directly maps to well understood concepts in principal-agent contexts. 

\textbf{What does adverse selection and moral hazard mean in the context of scheming?} The distinction is less clear than in traditional microeconomic models, largely because the notion of a ``contract'' is ill-defined (see e.g. \citet{hadfield2019incomplete}). In standard principal–agent settings, the distinction is clear: (a) a car seller holds private information about quality \textit{before} a sale, while (b) an employee may shirk \textit{after} being hired, when actions are unobservable. The former is adverse selection; the latter is moral hazard. This raises a natural question: what does it mean for human–LLM or LLM–LLM agents to enter a contract? While a full account is beyond the scope of this work, a natural interpretation is deployment outside an AI lab's security team (i.e. deployment to the public via an API). Under this view, an advanced LLM may conceal its type (e.g. aligned versus misaligned) before deployment, corresponding to adverse selection, while a deployed agent may acquire private information and withhold it from the principal, corresponding to moral hazard.



\section{Alternative views}
\label{sec:alternative}

We now take some time to address a selection of reasonable points of objection to certain aspects of the position taken in the work.

\textbf{(a)} \textit{Standard economic models assume utility-optimizers. Behavioral economics \citep{thaler1980} and prospect theory \citep{kahneman1979} challenge this notion. Why would we then expect LLM-based actors to be rational actors?}

Despite well-documented real-world deviations from utility maximization \citep{gennaioli2018}, rationality in economic models is a deliberate simplification that transforms complex behavior into tractable representations of incentive-driven decision-making, enabling external validity across heterogeneous settings \citep{raj2015, gibbons1997}. In multi-agent systems, LLM-based agents are explicitly trained and deployed to optimize objective functions, making utility-maximizing behavior a natural and often more appropriate modeling abstraction than in human settings \citep{ning2024, yang2020}. 

\textbf{(b)} \textit{Information asymmetry is not a fundamental problem with AI agents. This can be resolved with advances in technology, whereby AI systems can fully observe all other sub-agents}

We address this objection at two levels. First, we are not convinced that information asymmetry is non-fundamental. Language model agents store information in latent space as localized computations \citep{meng2022locating} and execute circuits tied to specific actions \citep{conmy2023towards}. It is therefore plausible that agents acquire task-relevant information through internal representations that are neither directly visible nor interpretable from model parameters. Techniques that explicitly move computation into latent space have already been proposed \citep{hao2024training}. Consequently, even if principals could observe internal states, this would not ensure reliable inference of agent intent, thereby inducing information asymmetry and making the principal-agent framework highly relevant. Second, even if information asymmetry were not ontologically fundamental, it is far more likely to arise and persist than not, since monitoring is costly and inefficient (see Sec. \ref{subsec:mechanism}), particularly as sub-agents receive task-specific inputs. Absent such costs, there would be little reason for delegation in the first place.

\textbf{(c)} \textit{Treating AI-based systems as principal-agent problems is misguided since we cannot design equally functioning mechanisms to address resulting inefficiencies}

Mechanism design explores the choice rules that induce agents to take costly actions that help achieve desired outcomes, even when the designer has incomplete information about agents' preferences. Consequently to reduce inefficiencies in multi-agent systems, mechanism design provides two very clear and concrete paths forward: (1) constrainig the space of feasible outcomes, and (2) design incentive schemes that motivate agents to undertake costly actions aligned with the principal's desired objective.

\section{How the research agenda should change}
\label{sec:research_agenda}

Multi-agent systems ought to have a separate \textbf{mechanism design agenda}. If scheming constitutes an equilibrium response for rational optimizing agents, we need research agenda that focuses on the design of mechanisms that reduce agency loss -- that is, induces agents to act on principal's behalf in such multi-agent systems. This raises the following questions \textbf{(a)} how to design multi-agent systems where agents reveal their capabilities (adverse selection)? What would such contracts or self-selection mechanisms look like? \textbf{(b)} How can we design multi-agent systems in which agents are incentivized to undertake costly actions that reliably align with the systems' designers objectives (i.e., address moral hazard)? \textbf{(c)} How can we enforce audit technologies which are triggered by output or behavior anomalies? \textbf{(d)} How can we adapt market-based mechanisms -- such as reputation and signaling systems \citep{spence1973, milgrom1982}, credit-like scoring \citep{stiglitz1981} and screening, \citep{wilson1977}, internal prediction markets \citep{wolfers2004}, tournament-style incentives \citep{lazear1981}, and other enforcement mechanisms from mechanism design -- to improve incentives and alignment in multi-agent systems?





\section{Discussion}
\textbf{Importance for AI safety}. There is little doubt that we are entering a phase in which autonomy is increasingly delegated to AI systems. Given the evidence to date, there is a strong reason to expect that such systems will face coordination problems. We argue that the core features of this problem, i.e. modeling AI agents under information asymmetry and goal-misalignment, are structurally equivalent to agency loss in the economic literature on mechanism design. Looking at this through the lens of principal-agent problems can help us understand why AI agents, implemented as language model policies under a delegation mechanisms, exhibit scheming behavior. The solution to such problems cannot end with custom reinforcement learning reward signals, as they have not stopped the emergence of self-preservation goals. Instead, we see this work as an urgent call to use the tools available in social science research to model multi-agent systems for what they are: principal-agent problems.

\textbf{Impact Statement}. We believe this work has impacts on at least several areas of machine learning, most notably AI safety, multi-agent interactions; and has the potential to shape how we design systems to develop safe and reliable AI.

\bibliography{example_paper}
\bibliographystyle{icml2026}

\newpage
\appendix
\onecolumn

\end{document}